\documentclass[english]{article}
\usepackage[T1]{fontenc}
\usepackage[latin9]{inputenc}
\usepackage{textcomp}
\usepackage{amstext}
\usepackage{amssymb}
\usepackage{graphicx}

\makeatletter

\DeclareRobustCommand{\greektext}{%
  \fontencoding{LGR}\selectfont\def\encodingdefault{LGR}}
\DeclareRobustCommand{\textgreek}[1]{\leavevmode{\greektext #1}}
\DeclareFontEncoding{LGR}{}{}
\DeclareTextSymbol{\~}{LGR}{126}
\newcommand{\lyxmathsym}[1]{\ifmmode\begingroup\def\b@ld{bold}
  \text{\ifx\math@version\b@ld\bfseries\fi#1}\endgroup\else#1\fi}

\makeatother

\usepackage{babel}
\begin{document}

\title{Self-assembly of anisotropic soft particles in two dimensions}

\author{Daniel Salgado-Blanco and Carlos I. Mendoza%
\thanks{E-mail: cmendoza@iim.unam.mx%
}\\
Instituto de Investigaciones en Materiales, Universidad Nacional\\
Autónoma de México, Apdo. Postal 70-360, 04510 México, D.F., Mexico}
\maketitle
\begin{abstract}
The self assembly of core-corona discs interacting via anisotropic
potentials is investigated using Monte Carlo computer simulations.
A minimal interaction potential that incorporates anisotropy in a
simple way is introduced. It consists in a core-corona architecture
in which the center of the core is shifted with respect to the center
of the corona. Anisotropy can thus be tuned by progressively shifting
the position of the core. Despite its simplicity, the system self
organize in a rich variety of structures including stripes, triangular
and rectangular lattices, and unusual plastic crystals. Our results
indicate that the amount of anisotropy does not alter the lattice
spacing and only influences the type of clustering (stripes, micells,
etc.) of the individual particles.
\end{abstract}

\section{Introduction}

Simple pair potentials are often used to describe effective interactions
among substances with supramolecular architecture \cite{denton,malescio3}.
Frequently, this is done considering models of particles consisting
of a hard core surrounded by a repulsive soft corona. In such representations,
the interactions between the so called core-shell or core-corona particles,
are treated in an effective way that takes into consideration the
entropy of the brush-like coronas surrounding each core. The simplest
of them consists of a hard core surrounded by a square shoulder \cite{malescio1,malescio2}.
Its simplicity allows to predict, based on geometrical arguments,
under which circumstances a given family of structures is obtained.
Among the physical systems described using this model we can mention
colloidal particles with block-copolymers grafted to their surface
where self-consistent field calculations lead to effective interactions
that can be modeled by a square-shoulder potential \cite{norizoe}.
Such interactions can be controlled by adjusting the length, species,
the grafting density of the grafted polymers, the quality of the solvent,
the density and location of the hydrogen bonding, etc.

The phase behavior and associated pattern formation of two-dimensional
systems with core-shell architecture has been studied by using numerical
simulations. These studies have shown that single component softened-core
repulsive potentials may give rise to stripe phases \cite{malescio1}-\cite{fornleitner}
and periodic structures that are explained in terms of the competing
interactions between the hard core and the soft shoulder. Binary mixtures
of cylindrically symmetric core-shell particles in two dimensions
in which particles of the same species interact through a repulsive
shoulder while particles of different species interact via an attractive
square well have also been studied \cite{mendoza}. It is found that
the larger number of control parameters inherent to the mixture allows
to extend the ensemble of possible self-assembled structures, originating
structures like stripes, square lattices as well as other more sophisticated
arrangements \cite{mendoza}. In spite of the rich variety of self-organized
structures found in the single and two-species systems of particles
with isotropic shapes and interactions, there are intrinsic limitations
associated with their isotropy \cite{lee}. In order to circumvent
such limitations, active research has been done in order to incorporate
anisotropy in the building blocks. Additionally, the unprecedent revolution
in particle syntheses, which has led to the possibility to modify
chemically the surface of the particles as well as to build an amazing
variety of building blocks of different shapes, compositions, patterns
and functionalities, offers a multitude of molecular designs for particles
with unique properties that are amenable to self-assembly \cite{vanakaras,glotzer 2007}.
A broad range of investigations has been focused on the fabrication
of non-spherical colloidal particles as well as their application
for drug delivery, molecular imaging and self-assembly (see \cite{lee}
and references therein). A variety of interesting and efficient techniques
to synthesize particles with different shapes have been devised in
recent years. Among them we can mention stretching methods, micromolding,
microfluidic approaches and methods using seeded polymerization \cite{lee}.
As a consequence, it is now recognized that particles with anisotropic
shape or interacting via anisotropic potentials are powerful tools
for engineering the assembly of particular targeted structure and
therefore have been actively investigated in recent years \cite{glotzer 2007,fejer}.
These types of colloidal particles have attracted increasing attention
because of their promising properties for applications in nanotechnology,
electronics, biotechnology and others. In particular, much work have
been done on rod-shaped particles \cite{veerman,vacha} two-faced
(Janus) \cite{hong}-\cite{jiang} and patchy particles \cite{glotzer 2004}-\cite{romano}. 

Recently, Vanakaras has proposed a special class of two-dimensional
Janus core-corona system that spontaneously form exotic and thermodynamically
stable structures not seen in systems of isotropic particles \cite{vanakaras}.
This and other studies, together with the large variety of experimental
techniques to fabricate core-shell and anisotropic particles suggest
that 2D anisotropic core-shell particles are promising candidates
for bottom-up design of precise two-dimensional templates.

Regardless of the myriad of possible shapes and interactions that
different anisotropy building blocks may have, all of them can be
conveniently classified in three different types \cite{fejer}, (1)
shape anisotropy + interaction isotropy; hard-rods belong to this
class, (2) shape anisotropy + interaction anisotropy, for example,
Gay-Berne \cite{gay} and patchy ellipsoids, and finally (3) shape
isotropy + interaction anisotropy, as in the case of spheres interacting
through a dipolar electric field. In what follows, we are going to
introduce a model for anisotropic particle of the type 3.

The aim of this paper is to propose a minimal model that incorporates
in a simple way the effects of anisotropy on the self assembly of
colloidal particles that interact via core-shell potentials. We study
by computer simulation the self-assembled structures obtained at zero
temperature for a collection of two-dimensional particles consisting
of a non concentric core-corona architecture. Despite its simplicity,
the rich pattern formation arising from the self organization indicates
the importance of the interaction anisotropy on these core-corona
systems. Among the rich variety of patterns encountered, we can highlight
the formation of stripes, rectangular arrays, triangular lattices,
and unusual plastic crystal phases. It is also found that anisotropy
may promote the stripe formation in some cases or inhibit their formation
in others, however, the degree of anisotropy does not alter the lattice
spacing. These results suggest a strategy for producing a range of
self-assembled structures which could be exploited for use as templating
or directing agents in materials syntheses.

\section{System and Methods}

The system consists of particles interacting through an anisotropic
pair potential $V(r)$ composed of an impenetrable core of diameter
$\sigma_{0}$ surrounded by an adjacent square shoulder with range
$\lambda\sigma_{0}$ similar to existing representations used to describe
isotropic branched particles \cite{malescio3}. In order to incorporate
interaction anisotropy in this model, we propose to shift the center
or the core with respect to the center of the corona as sketched in
Fig. 1. Mathematically, this potential can be expressed as
\begin{equation}
V(r)=\left\{ \begin{array}{l}
\infty,\text{ \ \ \ \ \ \ \ \ \ if core-core distance}\leq\sigma_{0}\\
\varepsilon,\text{ \ \ \ \ \ \ \ \ \ \ if corona-corona distance}\leq\lambda\sigma_{0}\\
0,\text{ \ \ \ \ \ \ \ \ \ \ otherwise}
\end{array}\right.,\label{potential}
\end{equation}
where $r$ represents the core-core distance for the infinite repulsive
part of the potential, and the corona-corona distance for the repulsive
square shoulder of energy $\varepsilon$. The only restriction for
the distance $d$ between the centers of the core and the corona of
a given particle is that the core remains completely contained inside
the corona, that is, $d\leq d_{max}$, where $d_{max}=\sigma_{0}\left(\lambda-1\right)/2$.
In order to maintain the interparticle potential as simple as possible,
we have assumed that there is no core-corona interaction. This model
could represent in an approximate way a branched particle in which
the length of the branches is larger in one direction than in the
opposite one. The parameter $\delta\equiv d/d_{max}$ is introduced
to quantify the displacement of the core with respect to the corona.
Because $0\leq d\leq d_{max}$, then the reduced shift $\delta$ assumes
values in the interval $0\leq\delta\leq1$. When $\delta=0$, the
shift is zero and the particle is a regular isotropic core-corona
disk. At the other extreme, when $\delta=1$, the shift is maximum
and the central disk representing the core is tangent from inside
to the disk representing the corona. Because of the different positions
of the core and corona, the cylindrical symmetry of the disk is broken,
imparting anisotropy to the particles.

Standard Monte Carlo simulations based on the canonical ensemble (NVT
simulations) in a square box of side $L$ with periodic boundary conditions
have been carried out using the Metropolis algorithm. We have used
$\sigma_{0}$ and $\varepsilon$ as length and energy units, respectively,
the reduced temperature $T^{*}=k_{B}T/\text{\textgreek{e}}$, where
$k_{B}$ is Boltzmann\textasciiacute{}s constant; the reduced number
density $\rho^{*}=N\sigma_{0}^{2}/L^{2}$ and the reduced displacement
of the center of the core from the center of the corona $\delta$,
as defined above. We have studied the pattern formation dependence
on $\delta$, for different choices of $\lambda$ and $\rho^{*}$
. Simulations are performed with $N=1000$ particles and control runs
with $N=5000$ particles to exclude finite size effects were also
done. In all cases, the system is first disordered at high temperature
and then brought from $T^{*}=0.4$ to the final temperature $T^{*}=0.01$
through an accurate annealing procedure with steps of $0.01$. An
equilibration cycle consisted, for each temperature, of at least $2\times10^{8}$
MC steps, each one representing one trial displacement and rotation
of each particle, on average. At every simulation step a particle
is picked at random and given a uniform random trial displacement
within a radius of $0.5\sigma_{0}$ and a random trial rotation within
an angle of $\pm0.1\pi$. If we are simulating inside a two-phase
coexistence region of the phase diagram, we can not exclude the possibility
that the simulation do not show the phase separation. This is due
to the fact that in NVT simulations the system may be kept at a density
where it would prefer to separate into two phases of different densities
but is prevented from doing so by finite-size effects. For this reason,
we have completed our investigation by studying the behavior of our
model system through simulations at constant number of particles,
pressure $P$, and temperature (NPT simulations). In this case, the
system is free to transform completely into the state of lower free
energy \cite{frenkel}.

A variety of interesting structural features of this simple model
at low temperatures are in evidence in Fig. 2, where a few of them
are exhibited for $\text{\textgreek{l}}=2.5$ $ $and $\text{\textgreek{r}\textasteriskcentered}=0.291$.
This set of parameters is chosen as to agree with the ones used by
Malescio and Pellicane in Ref. \cite{malescio3} to obtain lane formation.
Panel (a) shows a stripe pattern obtained for isotropic particles,
$\delta=0$. We observe that both, the interparticle distance in a
given lane and the lane to lane spacing is determined by the corona
diameter. The particle-particle distance in a given lane is such that
the coronas of each other particle just touch for the given set of
parameters. Analogously, the lane to lane distance is such that the
coronas of adjacent lanes barely touch. The system adopts this array
in order to minimize unfavorable overlaps between coronas. As we progressively
increase the anisotropy parameter, $\delta$, the system adopts alternative
structural motifs. For a small value of the reduced displacement,
$\delta=0$.2, the system organize in a pattern formed by two distinct
phases. The first one corresponds to the same lane pattern observed
in panel (a) but this phase co-exists with an approximately triangular
array of compact clusters or aggregates made mainly of dimers and
trimers. Particles that belong to a given cluster are located such
that their coronas overlap as much as possible. The limiting factor
that prevents the total collapse of such aggregates are their hard-cores
thus confering a non-cylindrically symmetric shape to the envelope
of the coronas of the resulting aggregates. For example, the aggregates
of trimers in Fig. 2b have triangular shapes as well as the envelope
of their coronas. On the other hand, the lattice parameter for the
lattice of clusters is such that the coronas of different clusters
do not overlap. This fact not only determine the lattice parameter
but, in complicity with the non cylindrically symmetric shape of the
cluster's coronas, also induces a orientational correlation between
different aggregates. Increasing even more the value of the reduced
displacement to $\delta=0.3$, the phase corresponding to the stripes
has disappeared completely and only an approximately crystalline arrangement
consisting of a triangular lattice of dimers and trimers remains.
Finally, at the maximum displacement, $\delta=1$, the regular lattice
of correlated dimers and trimers is replaced by an ordered array of
clusters, each one formed by two or three particles whose coronas
completely overlap but with the position of their cores completely
uncorrelated with the ones of the other aggregates. In other words,
the coronas that belong to an aggregate of particles overlap almost
perfectly forming an isotropic corona for the resultant aggregate.
This allows each aggregate to rotate freely and independently of the
other aggregates thus forming an unconventional plastic crystal with
long range (or quasi-long range) translational but no orientational
correlations. We say it is an unconventional plastic crystal since
usually the term refers to an array with translational order but without
correlations between the orientation of the molecules. Here, the lack
of orientational correlations is translated into a lack of positional
correlations between the cores. It is interesting to note that the
incorporation of anisotropy has inhibited the stripe formation for
this set of parameters.

Figure 3 shows a sequence of spatial configurations (top panels) for
$\text{\textgreek{l}}=3$ $ $and $\text{\textgreek{r}\textasteriskcentered}=0.5$.
As the relative displacement is increased, the system turns from the
staggered pattern of panel (a), where the (isotropic) particles are
arranged in quasiclump states with the center of the particles occupying
the lattice points of a subtriangular lattice (see inset), to a lane
conformation, for $\delta=0.166$, where each lane is constituted
by two adjacent rows of particles whose arrangement resemble a succession
of parallel dumbbells (Fig. 3b, top). The fact that the particle's
cores within a given stripe do not locate in a triangular sublattice
indicate that clustering is not close-packed in this case, in sharp
contrast with the case of isotropic potentials where clusters of close-packed
particles are predicted \cite{ziherl}. The spacing between consecutive
lanes is dictated by the condition that the coronas of adjacent lanes
do not overlap, as shown in the respective inset. With a further increase
of anisotropy ($\delta=0.333$), the system turns into a triangular
lattice of hexagonal clumps formed by seven particles whose cores
are located on a triangular sublattice, as shown in the inset of Fig.
3c (top). Finally, for the maximum anisotropy ($\delta=1$), the almost
perfect crystal lattice transforms into an unconventional plastic
crystal made of a regular lattice of discotic aggregates of mostly
five particles which are orientationally uncorrelated with the particles
forming other aggregates. Actually, in this case, the presence of
the cores do not influence the structure of the coronas, the same
triangular lattice of coronas would be obtained even if no cores were
present. Contrary to the case analyzed in Fig. 2, for this set of
parameters, the stripe formation is first promoted by the addition
of anisotropy (transition from Fig. 3a to Fig. 3b) and then suppressed
(transition from Fig. 3b to Fig. 3c). As seen, incorporation of anisotropy
has a profound effect in the final structure obtained.

We analyze the structural order of our system by looking at the diffraction
patterns (Fig. 3, central panels) and through the orientationally
averaged radial correlation function, $g(r)$, which gives the probability
to find a pair of molecules at a distance between $r$ and $r+dr$
(Fig. 3, bottom panels). In the case of isotropic particles (Fig.
3a), $g(r)$, is the same for the core and corona centers, as depicted
by the blue and red lines, respectively, of Fig. 3a bottom panel.
Clearly a long-range order is present with two large peaks, the first
one located near the core diameter $r\thickapprox\sigma_{0}$, showing
that the first coordination shell is formed by particles whose cores
are nearly touching, compatible with the snapshot of Fig. 3a. The
second large peak is located near the soft-shoulder diameter $r\thickapprox\lambda\sigma_{0}$,
showing that the outer lattice spacing is determined by the corona
diameter. Intermediate peaks located near $r\thickapprox2\sigma_{0}$,
indicates that the second coordination shell consists of particles
located at twice the core diameter, consistent with the fact that
the underlying triangular lattice is not completely filled, otherwise,
peaks located near $r\thickapprox\sqrt{3}\sigma_{0}$, would be present.
In Fig. 3b bottom panel we show $g(r)$, for a small anisotropy, $\delta=0.166$.
Since the center of the cores do not coincide with the center of the
coronas, then there will be a $g(r)$ for the cores and a different
one for the coronas. Although $g(r)$ for the cores still present
moderate long-range correlations (blue line), its peak structure is
less pronounced with the exception of the first peak, as a result
of a slightly more disordered arrangement. In contrast, the peak structure
of $g(r)$ for the coronas (red line) is very pronounced indicating
the presence of long-range order. Notice that the first peak appears
below $r=\sigma_{0}$ due to the fact that the relative displacement
$\delta>0$ allows the center of the coronas to approach closer. The
large peak around $r\thickapprox\lambda\sigma_{0}$ is the signature
of the interlane spacing which is, again, imposed by the shoulder
width. At the next case considered, $\delta=0.333$, the shape of
$g(r)$ still presents considerable structure as shown in Fig. 3c,
bottom panel (blue line). The first coordination shell is well pronounced
and the second and third peaks, near $r\thickapprox\sqrt{3}\sigma_{0}$
and $r\thickapprox2\sigma_{0}$, correspond to the second and third
nearest neighbors. The broader peaks in $g(r)$ represent correlations
between particles belonging to different aggregates. The larger anisotropy
of this case ($\delta=0.333$), allows a larger interpenetration of
the coronas belonging to the same aggregate, a fact which is reflected
by the first broad structured peak in $g(r)$ for the coronas, as
shown in Fig. 3c bottom panel (red line). The second broad peak appearing
near $r\thickapprox\lambda\sigma_{0}$, corresponds to the first coordination
shell of the triangular lattice formed by the aggregates. Finally,
in Fig. 3d, bottom panel we show $g(r)$ for the largest anisotropy
($\delta=1$). An oscillatory form, typical of a liquid-like structure,
is seen in the case of the cores (blue line) with only a few correlation
layers. On the other hand, the triangular lattice of overlapped coronas
with lattice constant $\lambda\sigma_{0}$ is clearly reflected by
the peaks at $r\thickapprox0$ and $r\thickapprox\lambda\sigma_{0}$
(red line). Therefore, in this last case, the liquid-like structure
of the cores coexist with a triangular lattice of overlapped coronas.
This characteristic suggest the possibility of using this self-assembled
structure as a template to tailor materials with hybrid properties
at the nanoscale, those of an amorphous material (associated with
the cores) mixed with those of a crystalline material (associated
with the coronas).

We further analyze the structural order of our system through the
diffraction patterns (Fourier transform). The phases' rotational symmetry
as well as the micelles' structure is revealed by these patterns as
shown in the central panels of Fig. 3 for both, the cores (left) and
the coronas (right). In the case of isotropic particles, $\delta=0$,
both patterns coincide and show the triangular symmetry of the underlying
lattice as well as the peculiar staggered distribution of the particles
on it. For the stripe pattern shown in Fig. 3b, where $\delta=0.166$,
the diffraction patterns for the cores (Fig. 3b central, left) is
slightly different from the corresponding pattern for the coronas
(Fig. 3b central, right) but in both cases the lane geometry is shown
and not underlying triangular lattice is present, in contrast to the
other cases and in accordance with the inset of Fig. 3b upper panel.
As we increase the anisotropy to $\delta=0.333$, the distinction
between the diffraction patterns for the cores (Fig. 3c central, left)
and the coronas (Fig. 3c central, right) is more clearly visible.
In the first case, the internal triangular arrangement of the micelles
is evident, while for the coronas this is not the case since they
almost completely overlap in any given micelle. Finally, for the maximum
anisotropy, $\delta=1$, the disordered arrangement of the cores produces
a diffuse ring in the diffraction pattern (Fig. 3d central, left)
while the pattern for the coronas shows the signature of a triangular
lattice (Fig. 3d central, right).

From the above results, it appears that there are two length scales
associated with the pattern formation, one associated with the underlying
sublattice where the cores are located (except in the case of the
plastic crystal, where the core positions do not lie in a lattice)
and other, associated with the coarse grained lattice or stripe patterns,
where the clumps or the lanes are located. After the analysis of the
radial distribution function, it is clear that this length scales
are associated with the intrinsic length scales of the particles,
namely, the core and corona diameters, $\sigma_{0}$ and $\lambda\sigma_{0}$,
respectively. However, there is a third length intrinsic to the particles
which is the relative displacement between the core and the corona,
$\delta$. Our results show that this length does not intervene in
the lattice spacing but rather in the type of lattice formed: stripes,
crystal, or plastic crystal. The number of particles that form the
clumps or the width of a given lane are determined by the reduced
density of the sample, $\text{\textgreek{r}\textasteriskcentered}$,
and also by the shoulder width.

Two final representative patterns formed by the present model are
shown in Fig. 4. Here, a zigzag pattern is shown in panel (a) for
$\text{\textgreek{l}}=2$, $\text{\textgreek{r}\textasteriskcentered}=0.6$
and $\delta=0.5$; and a rectangular lattice of cores is shown in
panel (b) for $\text{\textgreek{l}}=2$ and $\delta=0.34$. The last
example shown is obtained by means of a NPT simulation in which the
pressure is chosen so that the density is $\text{\textgreek{r}\textasteriskcentered}=0.7$.
Also, the shape of the simulation box is allowed to change to disregard
a possible influence of the simulation box geometry on the final arrangement.
Due to the simplicity of the potential it is possible to extract the
lattice constants of this arrangement. The shortest axis, $a$, is
determined by the condition $a=\sigma_{1}$ while $b=\left(\sqrt{3}/2\right)\sigma_{1}+\sigma_{0}/2-d$,
and this lattice can only appear if $\lambda\geq\sqrt{3}$. For the
parameters of Fig. 4b, $a=2$ and $b=2.06$.

\section{Conclusions}

In conclusion, we studied the self assembly of core-corona particles
interacting via an anisotropic potential in two dimensions. The model
contains a minimal number of geometrical parameters and despite its
simplicity, it exhibits a rich variety of patterns. Among them we
can highlight the formation of lanes, triangular and rectangular lattices,
and unusual plastic crystals with long-range translational correlations
for the coronas but with liquid-like structural features for the cores.
This opens up the fascinating possibility to design an hybrid self
assembled material at the nanoscale with properties that share some
aspects of an amorphous and other aspects of a crystalline material.
Our results indicate that the amount of anisotropy does not alter
the lattice spacing and only influences the type of clustering (stripes,
micells, etc.) of the individual particles. We believe that the model
investigated in this work, although highly simplified contains the
main ingredients to understand the self organization of discotic soft
particles in 2D with anisotropic interactions.

{\LARGE Acknowledgements}{\LARGE \par}

This work was supported in part by grant DGAPA IN-115010.

\newpage{}

{\LARGE Figure Captions}\bigskip{}

Fig. 1. Schematics of the model. Pair potential $V(r)$ as a function
of the distance $r$ between two particles interacting via an off-centered
core-corona architecture. $r$ corresponds to the core-core distance
for the hard core (blue line) and to the corona-corona distance for
the square potential (red line). The upper right corner depicts the
anisotropic particle. The filled blue circle represents the hard core
and the red circunference is the external edge of the soft corona.

Fig. 2. Spatial arrangements of the system at temperature $T^{*}=0.01$
for $\lyxmathsym{\textgreek{l}}=2.5$ and $\lyxmathsym{\textgreek{r}}^{*}=0.291$
(NVT simulation). (a) Stripe pattern for $\delta=0$, (b) stripes
mixed with a crystalline array made mainly of dimers and trimers obtained
for $\delta=0$.2, (c) crystalline array of dimers and trimers for
$\delta=0.3$, (d) plastic crystal made of a triangular array of aggregates
formed mainly by two and three particles whose coronas overlap completely
and can rotate independently of the other aggregates for $\delta=1$.

Fig. 3. Structural properties at $T^{*}=0.01$ for $\lyxmathsym{\textgreek{l}}=3$
and $\lyxmathsym{\textgreek{r}}^{*}=0.5$ (NVT simulation). Upper
panels: Snapshots of a sequence of configurations. (a) Wigled pattern
for $\delta=0$, (b) stripes for $\delta=0.166$, (c) triangular lattice
of heptamers for $\delta=0.333$, and (d) plastic crystal made of
a triangular array of aggregates formed mainly by two and three particles
whose coronas overlap completely for $\delta=1$. Central panels:
Diffraction patterns for the cores (left) and coronas (right) for
the snapshots shown in the corresponding upper panels. Bottom panels:
Radial distribution function, $g(r)$, for the cores (blue line) and
coronas (red line) for the configurations shown in the respective
upper panels.

Fig. 4. Spatial arrangements of the system at temperature $T^{*}=0.01$
for $\lyxmathsym{\textgreek{l}}=2$. (a) A zigzag pattern obtained
for $\text{\textgreek{r}\textasteriskcentered}=0.6$ and $\delta=0.5$
(NVT simulation). (b) Rectangular lattice obtained for a density $\lyxmathsym{\textgreek{r}}^{*}=0.7$
and $\delta=0.34$ (NPT simulation). The lines are guides for the
eye to show the rectangular lattice whose parameters are $a=2$ and
$b=2.06$.

\newpage{}
\begin{figure}
\includegraphics[scale=0.75]{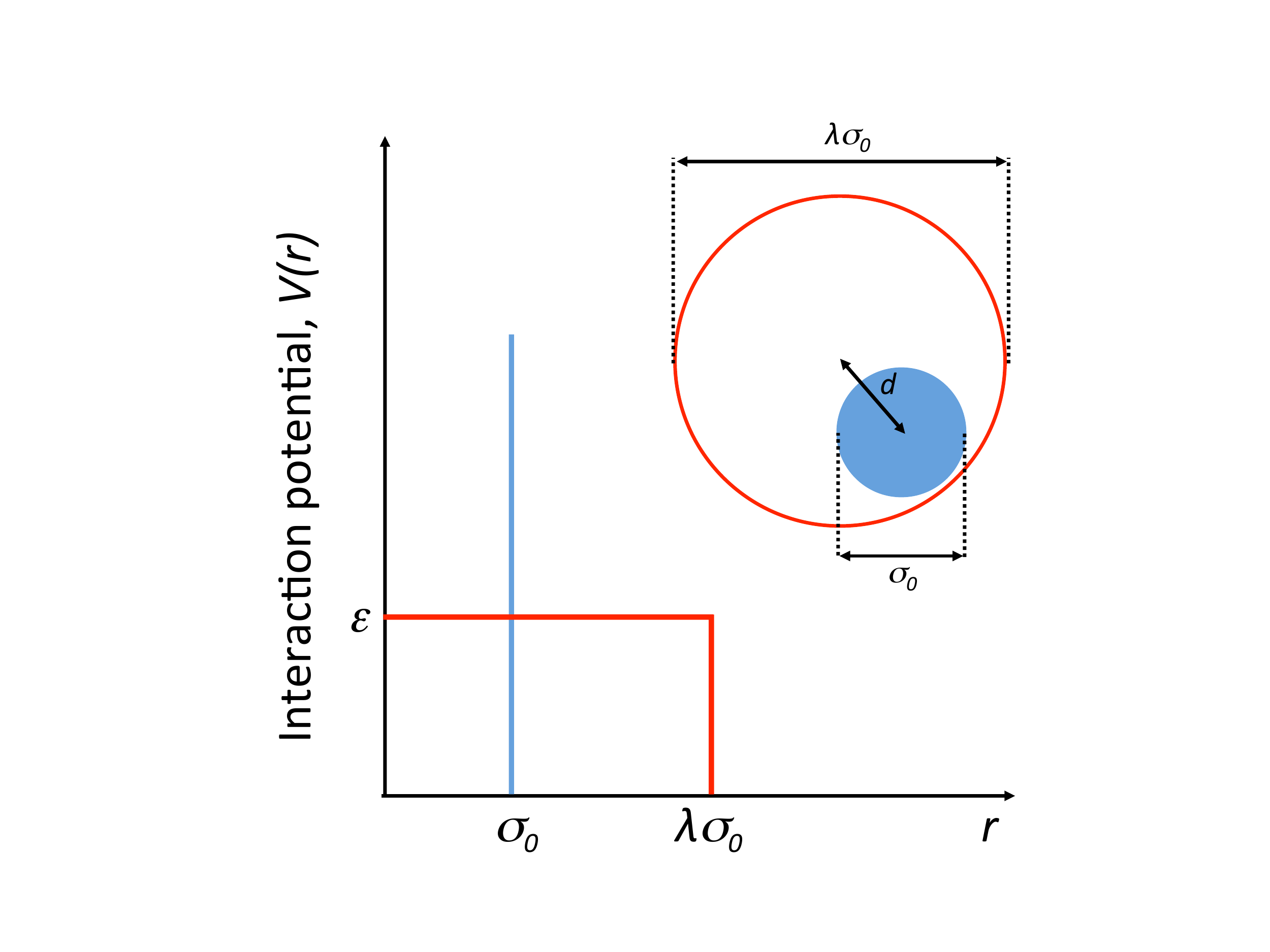}

\caption{Schematics of the model. Pair potential $V(r)$ as a function of the
distance $r$ between two particles interacting via an off-centered
core-corona architecture. $r$ corresponds to the core-core distance
for the hard core (blue line) and to the corona-corona distance for
the square potential (red line). The upper right corner depicts the
anisotropic particle. The filled blue circle represents the hard core
and the red circunference is the external edge of the soft corona.}

\end{figure}

\begin{figure}
\includegraphics[scale=0.75]{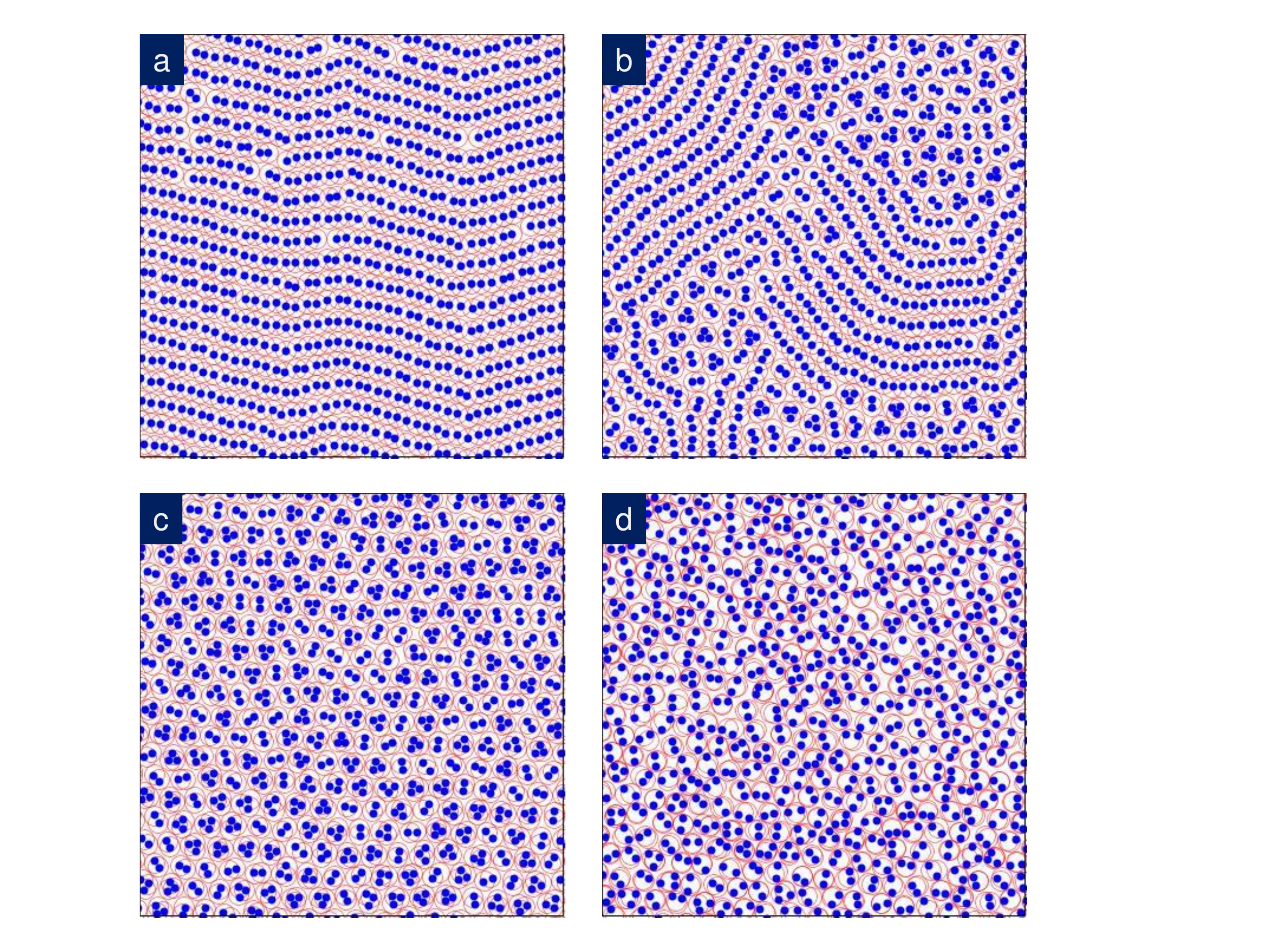}\caption{Spatial arrangements of the system at temperature $T^{*}=0.01$ for
$\lyxmathsym{\textgreek{l}}=2.5$ and $\lyxmathsym{\textgreek{r}}^{*}=0.291$
(NVT simulation). (a) Stripe pattern for $\delta=0$, (b) stripes
mixed with a crystalline array made mainly of dimers and trimers obtained
for $\delta=0$.2, (c) crystalline array of dimers and trimers for
$\delta=0.3$, (d) plastic crystal made of a triangular array of aggregates
formed mainly by two and three particles whose coronas overlap completely
and can rotate independently of the other aggregates for $\delta=1$.}
\end{figure}

\begin{figure}
\includegraphics[scale=0.5]{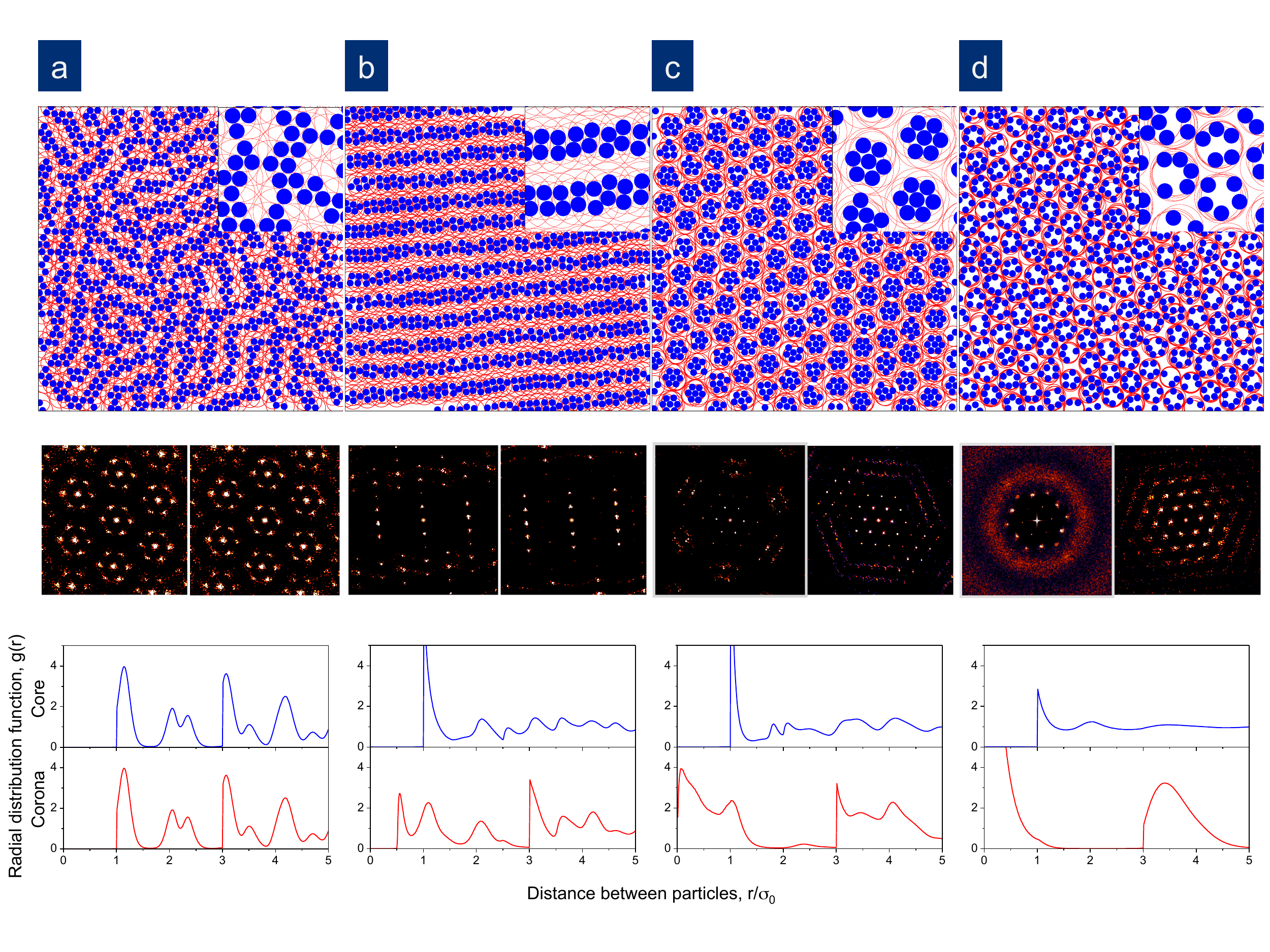}\caption{Structural properties at $T^{*}=0.01$ for $\lyxmathsym{\textgreek{l}}=3$
and $\lyxmathsym{\textgreek{r}}^{*}=0.5$ (NVT simulation). Upper
panels: Snapshots of a sequence of configurations. (a) Wigled pattern
for $\delta=0$, (b) stripes for $\delta=0.166$, (c) triangular lattice
of heptamers for $\delta=0.333$, and (d) plastic crystal made of
a triangular array of aggregates formed mainly by two and three particles
whose coronas overlap completely for $\delta=1$. Central panels:
Diffraction patterns for the cores (left) and coronas (right) for
the snapshots shown in the corresponding upper panels. Bottom panels:
Radial distribution function, $g(r)$, for the cores (blue line) and
coronas (red line) for the configurations shown in the respective
upper panels.}

\end{figure}

\begin{figure}
\includegraphics[scale=0.6]{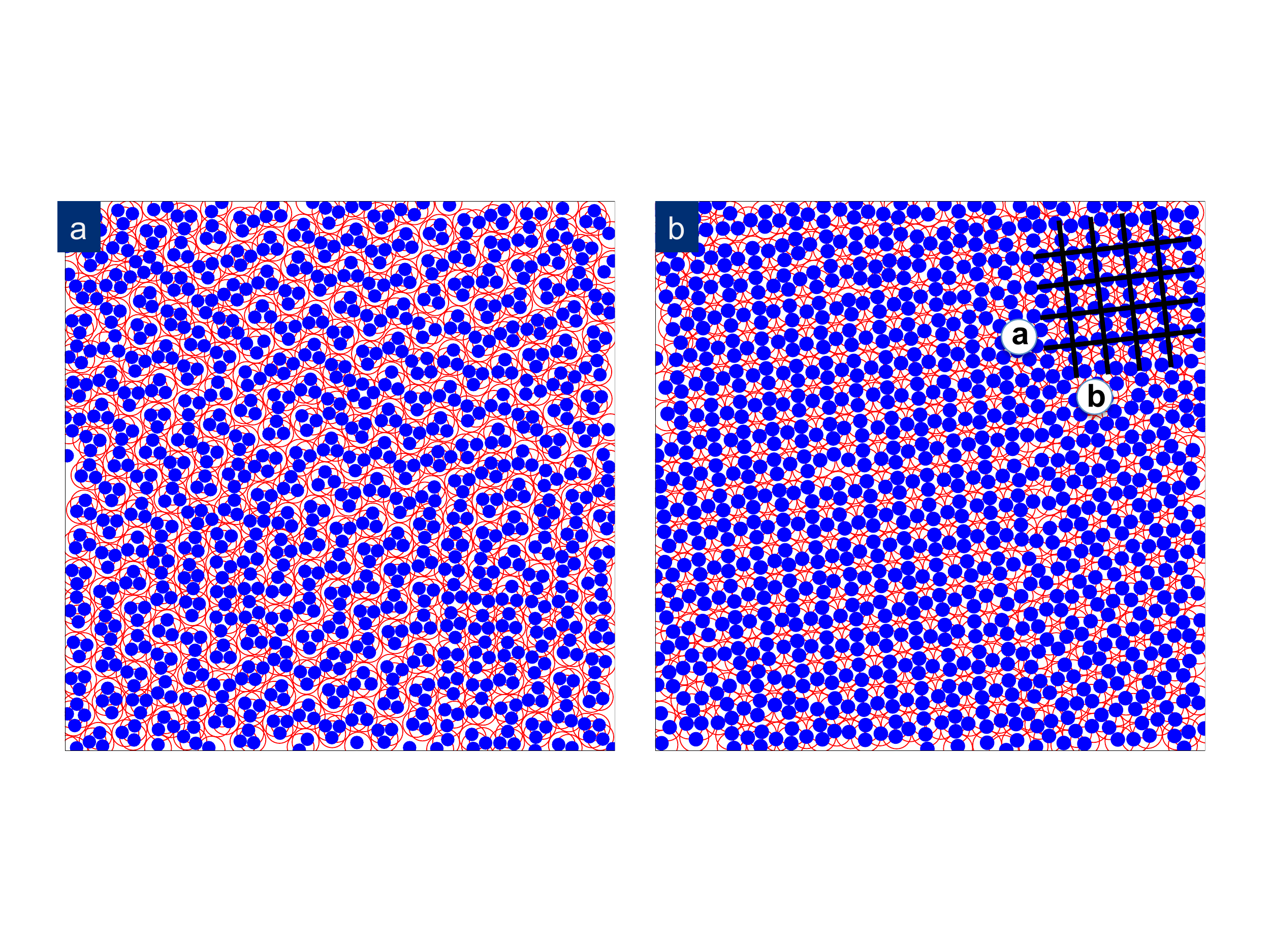}\caption{Spatial arrangements of the system at temperature $T^{*}=0.01$ for
$\lyxmathsym{\textgreek{l}}=2$. (a) A zigzag pattern obtained for
$\text{\textgreek{r}\textasteriskcentered}=0.6$ and $\delta=0.5$
(NVT simulation). (b) Rectangular lattice obtained for a density $\lyxmathsym{\textgreek{r}}^{*}=0.7$
and $\delta=0.34$ (NPT simulation). The lines are guides for the
eye to show the rectangular lattice whose parameters are $a=2$ and
$b=2.06$.}
\end{figure}

\end{document}